# Prichain II: CloudGuardian
# Cloud Security Proposal with Blockchain

R. C. Rodrigues , P. M. C. Mateus  and V. R. Q. Leithardt (Senior Member, IEEE)

*Abstract* — **With the advancement of cloud computing, data storage, and security have become crucial. The growing adoption of cloud services by companies, accompanied by increased threats from cybersecurity, highlights the importance of privacy and ownership of user data. Between 2022 and 2023, there has been an increase of around 48% in cloud security threats, emphasizing the urgent need for strong security solutions. To face these challenges, in this project, we propose integrating the Ethereum network's blockchain technology with a database located in the PostgreSQL cloud. The proposed solution aims to provide bidirectional data synchronization and strict control of access mechanisms. Blockchain technology ensures immutability and transparency of transactions, while PostgreSQL provides efficient and scalable storage. Through rigorous testing in an adaptive traffic control scenario, the results obtained indicate that this solution offers a significantly high level of security due to the decentralization of data, confirming that this solution is effective, and making it a powerful new option to improve security in cloud environments. In conclusion, the solution proposed in this project not only increases information security but also demonstrates the practical feasibility of integrating blockchain with cloud relational databases. This two-way alignment improves protection against cyberattacks and ensures that user data is protected from unauthorized access and malicious changes.**

## I. INTRODUCTION

The project aims to combine cloud databases with blockchain technology to store and protect information (in cloud computing). The main idea is to connect PostgreSQL databases to the Ethereum blockchain to guarantee better data integrity and access control mechanisms. By taking this route, we can have decentralized storage along with clear (or unambiguous) data transmission, which allows people to have full control over their data while maintaining its confidentiality. The report describes the methods used, outlines the methodology, including smart contract development, backend and frontend application creation, permission logic implementation, and testing procedures.

Its major objectives include establishing a durable system that enables the blockchain network to interact with cloud databases, putting in place access control mechanisms for safeguarding user rights to privacy as well as authentication and authorization provisions so that the integrity of information could be maintained due to such things like unchangeable nature of this technology. We will design a user-friendly web interface with options to log in or register for an account and features that allow access to and update information. There will be a testing phase to check if the project is feasible and evaluate the efficiency of algorithms used in various parts of the project, besides identifying ways for it to be made better.

This project was inspired by the increasing use of cloud computing for data storage and management, along with the growing cybersecurity threats that organizations face. As a result, many organizations have shifted their activities to cloud technology due to its associated benefits, such as affordability and accessibility. However, cloud databases encounter significant challenges related to security, compliance, and data privacy. To address these challenges, this project aims to create a secure and transparent way to store data by combining blockchain technology with cloud database systems. The immutable and decentralized nature of blockchain will help ensure data integrity, improve access control, and reduce the risk of tampering or unauthorized access.

## II. RELATED WORK

The articles below have all studied the interaction between blockchain and cloud computing but have some differences between them.

According to the scientific article [8], the authors explain the fundamentals of blockchain technology, including blocks, hashes, and ledgers, and describe the challenges of authentication in cloud computing.

According to article [9], it is intended to correct the privacy problem by introducing a new access control method, AuthPrivacyChain. This new method uses smart contracts to manage access permissions and encryption to ensure data privacy.

According to the scientific article, the authors Filecoin, Sia, and Storj [10], compare cryptographic methods such as lightning networks to enable efficient transactions, offering greater reliability and resource optimization.

The study [11], focuses on using blockchain with advanced encryption. It collects user data, generates SHA-256 digital signatures, and encrypts data using the elliptic curve Diffie-Hellman scheme. It inspired us to encrypt sensitive data in our project to increase the security of our project.

Table 1 presents a comparison between the works and ours regarding some features. It highlights that none of the work integrates all the functions proposed in this work. Furthermore, it is worth noting that the resources provided in some of the commercial solutions are not restricted to free versions, such as in this case the cloud database and the Ethereum test network. In addition to the related works compared and presented in table 1, other scenarios and applications are based on the works developed in [23-28]. These works present scenarios and discussions as well as descriptions, criteria, parameters and concrete definitions of investigations carried out on the topic and area, which require added value related to Blockchain and data management.

**Table I. Comparison of related works**

| Study | Assets | Storage Method | Blockchain Plataform | Blockchain Type |
|---|---|---|---|---|
| Application of Block chain in Cloud Computing | Data storage and database transactions | Blockchain | Not specified | Public |
| AuthPrivacyChain: A Blockchain-Based Access Control Framework with Privacy Protection in Cloud | Cloud and data resources | Blockchain | EOS (Enterprise Operation System) | Public |
| Blockchain-based Decentralized Storage Scheme | User data | Blockchain | Ethereum | Public |
| Blockchain assisted encryption | User details, digital signatures and encrypted data | Audio and video files in Inter Planetary File System | Not specified | - |
| CloudGuardian - Cloud Security Proposal with Blockchain | Data storage, database transactions, cloud and user data | Relational Database and Blockchain | Ethereum (Sepolia Testnet) | Public |

## III. TECHNOLOGIES AND TOOLS

### A. PostgreSQL Coud Database

The decision for our cloud database is an essential point in this project, and the decision was between MySQL and PostgreSQL, as they stand out as the two most popular database systems available today. PostgreSQL is known for being the best in data integrity, with ACID-compliant transaction support that is essential for keeping consistency and reliability in place However, after reading this article [18], we opted for PostgreSQL, where we resorted to creating the cloud database using ElephantSQL.

### B. Ethereum Sepolia Network

According to article [21], the Ethereum blockchain concept can be seen as a state machine that executes based on transactions that exist within it (page 3). A test network named Sepolia exists in addition to the main Ethereum network. Sepolia has been designed for test and development purposes to ensure that it does not have any effect on the main Ethereum network. It is a platform that enables developers to try out smart contracts, decentralized apps as well as other blockchain features without any risk involved. It was this network that we used in our project. The supplier chosen to use the Blockchain network was Alchemy, due to a series of reasons that make this platform highly advantageous for the project in question. Alchemy is recognized as a leading infrastructure provider for decentralized applications (dApps) and development on the Ethereum blockchain.

### C. Smart Contracts

Smart contract technology, embedded in blockchains, is revolutionizing conventional industries and business processes by automatically enforcing contractual terms without the need for a trusted third party. This automation reduces administrative and service costs, improves efficiency, and minimizes risks [22]. In the project, smart contracts are implemented on the Sepolia testnet, belonging to Ethereum, where they are written in the Solidity programming language.

These are strategically deployed to automate data transactions between the blockchain network and cloud databases, ensuring that predefined conditions are met before storing or updating data. The insertion of smart contracts into the blockchain network is carried out through a deployment process, which involves compiling the "insert_contract" contract source code and sending a special transaction to the Ethereum network to deploy the contract. Once deployed, we return the contract address so we can interact with it. It is possible to observe in Fig. 1, the source code for implementing the contract on the Ethereum network

```
const {ethers} = require("hardhat");
async function main(){
    const optimized_data = await
ethers.getContractFactory("DataOptimized");
    const od = await optimized_data.deploy();
    console.log("O contrato foi inserido no endereço: ", od.address);
}
main()
    .then(() => process.exit(0))
    .catch(error => {
      console.error(error);
      process.exit(1);
    });
```

**Fig. 1 Contract source code to deploy the contract**

A contract called "DataStorage" was created with a data structure to store vehicle information, such as the registration number and ID of the respective device that captured it. This contract consequently includes methods to interact with this data to store new data on the blockchain and on the other hand, another function to be responsible for retrieving this information securely and transparently. When deployed on the blockchain, the smart contract becomes immutable and can be invoked by any user to perform specific operations.

The use of events allows other interested parties in the network to be notified whenever new data is stored, thus allowing several advantages such as the impossibility of falsifying data on the network. It is possible to observe in Fig. 2, the source code developed for the smart contract that allows interaction between the Ethereum network and the project system.

```
pragma solidity >=0.8.19;
contract DataOptimized {
    struct Dataoptimized{
        string allData;
    }
    mapping (string => Dataoptimized) public data;
    event dataStored(string id, string str);
    event dataRemoved(string id);
    function storeData(string memory id, string memory str) public {
        data[id].allData = str;
        emit dataStored(id, str);
    }
    function getData(string memory id) public view returns (string memory allData){
        return data[id].allData;
    }
    function removeData(string memory id) public {
        delete data[id];
        emit dataRemoved(id);
    }
}
```

Fig. 2 Source code for the smart contract

### D. Hardhat

Hardhat is an essential tool for developing smart contracts on the Ethereum platform, which has several functions that help create and deploy smart contracts on the network. It plays a key role in the smart contract development lifecycle, from writing the code in Solidity language to its deployment and maintenance. Without this tool, we cannot interact with the smart contract.

### E. MetaMask

MetaMask is an encrypted digital wallet that lets you manage your accounts, keys, and tokens securely. For developers, MetaMask provides access to Ethereum's API. This allows for transaction confirmations and cost indications using ETH (Ethereum's currency) to make the transactions into the node. At the time of project development, Sepolia ETH can be obtained from the Sepolia testnet faucet, which allows anyone to send a small amount of Sepolia ETH to the Testnet network for their wallet. Sepolia ETH is the currency used to pay the price of transactions on the Sepolia testnet, like how ETH is used to pay for computation on the Ethereum mainnet, which is also a proof-of-work chain before the merge. A pop-up is displayed indicating that the transaction was sent successfully, with the amount of Sepolia ETH deposited in the wallet.

### F. Express.js

Express.js is a framework for Node.js that simplifies the creation of web applications and APIs. In the project, Express.js was used to create a web server that manages endpoints to interact with the database and the Ethereum blockchain and perform operations such as sending and receiving data, all concisely and efficiently. It was applied for authentication, user registration, interaction with a smart contract on Ethereum, and data manipulation in the PostgreSQL database.

## IV. SOLUTION ARCHITECTURE

The system is designed to securely store confidential data, such as in this project's scenario, vehicle information, in a PostgreSQL-type database hosted in the cloud. This will require backend programming to ensure smooth interaction between the database and smart contracts deployed on the Ethereum blockchain. The Ethereum blockchain has a testnet called Sepolia for blockchain technology. To make it easier for us to deploy and manage our blockchain networks, we use the Alchemy provided by the blockchain as a service (BaaS). The Ethereum blockchain employs Proof of Stake (PoS) consensus algorithms and introduces smart contracts, which automate processes based on predefined conditions, ensuring transparency and integrity within the system. It is possible to observe in Figure 3, the prototype proposal, in which it is intended to merge blockchain technologies and cloud databases.

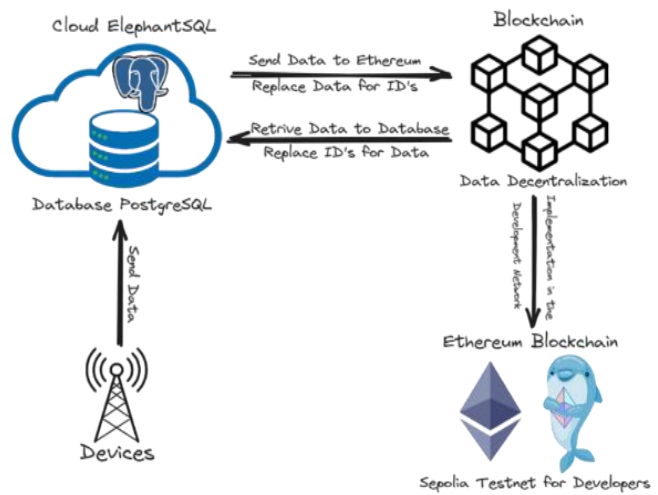

**Fig. 3 Architecture proposal**

To ensure a cohesive and visually appealing implementation of the API, it was crucial to first develop an initial layout proposal for the web. The Front-End application has a connection to the API Gateway via HTTP requests, presenting information in pages for viewing and managing data.

The layout proposal covered key elements of the API, such as the Login, Registration, Menu, Interaction with Database, and Retrieve Data from Ethereum pages. Figure 4 below presents a flowchart that represents the general functioning of our system based on user management and data management. It describes the decisions and conditional paths that determine execution based on user input, and includes time management elements, allowing for scheduled tasks to happen periodically.

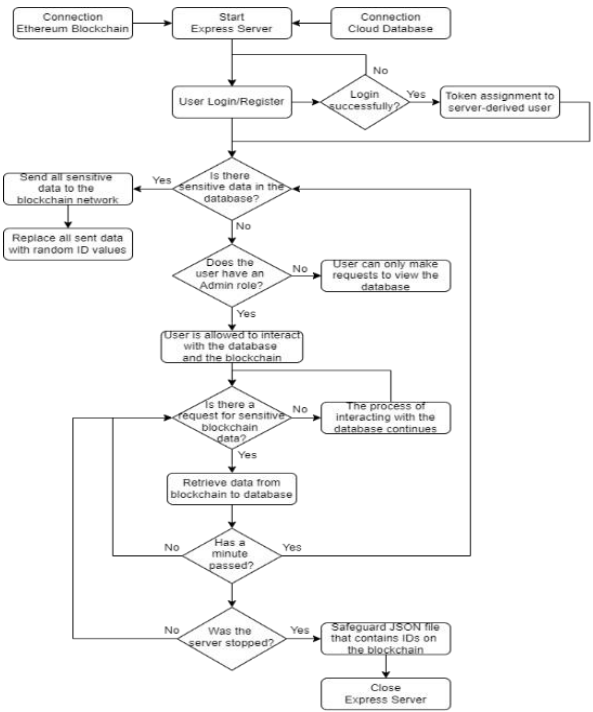

**Fig. 4 Flowchart of architecture proposal**

## V. IMPLEMENTATION

### A. First Preliminary Version

In the first development version of the system, the minimum requirements were long, and the focus was only on the correct functioning of the system. In other words, obtaining each data package is composed of sensitive information, in this scenario, only a vehicle registration and an ID of the device that captured it, are present in the database. Each packet is sent to the blockchain, where it must be extracted and confirmed to complete the blockchain transaction process. In this process, each package is associated with an ID, which is inserted into the database replacing the real values that are sent to the blockchain along with this associated ID. This association of an ID with real values is of great importance when there is a need to retrieve data back to the database. After developing these algorithms and running the system, two major problems were found:
1. Large latency corresponding to mining and transaction confirmation for each packet.
2. Low security of data relating to assigned IDs that are present in the database and are the same present in the blockchain.

The ID's security problem lies in its exposure, as it is present in the database, and if a malicious person gained access to the smart contract address, they would have the possibility of recovering the real data on the blockchain. Therefore, after observing the low efficiency and security of the system, it was necessary to develop the optimization of this version of the system.

### B. First Optimized Version

To solve the problems presented in the first preliminary version of the system, a new solution optimized for data processing and security was designed. The mechanism for obtaining all the data present in the database individually, and subsequently sending it to the blockchain, had to be changed. To do this, we use an array, which represents a single package with all this data currently existing, and then use a Map, where the package is associated with a generated ID value that will be sent to the blockchain. In the Map, the ID that will be sent to the blockchain will correspond to the Key and the array/package will correspond to the value associated with this Key. With this new data processing method, it was possible to optimize the transaction process, as only a single package was sent to the blockchain that contains all the data currently present in the database instead of the data individually. In this way, less pressure was placed on the blockchain, reducing the large amount of data input that is directly related to the number of transactions and consequently reducing the latency previously existing in the data transaction for the blockchain and the database. In this way, the method of using IDs sent to the blockchain was replaced, in which in the first preliminary version of the system they were the same as those in the database. Thus, new values of randomly generated IDs were created to be placed in the database, where these new random IDs present in the database do not directly correspond to any ID present in the blockchain, where these are only found in the Map, which is in process internal.

However, it is important to note that this solution introduces a compromise in terms of performance and security, as it performs several transactions to retrieve all existing data on the blockchain for the database:
1. Increased processing and latency when processing many transactions.
2. Reduced security by making all sensitive data visible after all transactions carried out.

### C. Second Proposed Version

In addition to strengthening security measures, a new approach has been developed to resolve inefficiencies found during the process of retrieving data from the blockchain to the database. In the previous iteration of the system, retrieving data from the blockchain required removing the corresponding data entries from the database, a process that could incur significant time overhead, especially for many upload transactions to the blockchain. To overcome this challenge, a new solution was proposed in which, instead of directly transferring the blockchain data back to the database, it is possible to download a JSON file containing the limiting information regarding the customer's desire. Thus, the customer can request data relating to a specific time interval or to determine which vehicles were captured by a certain device (reference to a geographic location), it only presents pertinent/limited data in the downloaded file instead of all the data that exist on the blockchain. By bypassing the need to remove individual transactions from the database, this approach speeds up the data retrieval process, completely removing the time required to retrieve data between the blockchain and the database. In this way, there are only transaction processes between the database and the blockchain when sending data.

However, it is important to note that this solution introduces a compromise in terms of security, leaving the system potentially vulnerable to certain security threats:
1. Does not require the generation of new IDs that were created whenever sensitive data was introduced or returned by the blockchain to the database.

*D. Third Proposed Version*

To solve the problems presented in the first optimized version of the system, a new solution was designed, this being the third version of the system, this one to optimize data processing. The data retrieval mechanism used in the first optimized version, remembering that it sent all the data present in the database in a single package, and subsequently sent to the blockchain, remained unchanged, and it was desired to change the data retrieval mechanism of the blockchain. To optimize data processing in the recovery transaction, that is, data coming from the blockchain to the database, it was necessary to implement some filters/restrictions in this process. To this end, it was decided to implement a system that allows the customer to request data relating to a specific time interval or day to determine which vehicles were captured by a specific device (reference to a geographic location). Thus, transaction processes will be significantly reduced, and instead of processing all existing data on the blockchain, this process ends up being limited. In this way, it was possible to cause less pressure on the system, significantly reducing the large amount of data processing and consequently reducing the latency previously existing in the first optimized version with the transaction of all data from the blockchain to the database. On the other hand, the security of the system was also consequently increased as it prevented visibility of all data in the database that was on the blockchain to all customers after a recovery process at a given time. Given this, in this new solution, only pertinent data is presented in the database.

*E. Improvements in all proposed versions*

There was a curiosity to look for more improvement solutions that aimed to strengthen security and optimize data recovery processes.

First, a security vulnerability was identified regarding the locally stored "id_mapping.json" file. This file was stored in plain text format, making it susceptible to unauthorized access in the event of a server compromise. This must be shared in the distributed system, essential to maintain the integrity and consistency of data in cases of server failure, and to this end, the global operation of the system must not be compromised by running it on another machine belonging to the distributed system. This method increases the scalability, reliability, and availability of the system. To mitigate the risk of file exposure, the bcrypt library ensures that the contents of the "id_mapping.json" file always remain secure.

By encrypting the file, even if unauthorized access to the server occurs, the confidential ID values stored in it remain protected from prying eyes, thus strengthening the overall security posture of the system. A strong encryption mechanism was implemented using the It is also important to mention that the database connection process was also optimized, using Connection-Pools, which instead of opening and closing a connection whenever necessary, the program maintains a set of open database connections that are reusable as needed, which is an efficient practice for handling database connections in applications that contain data concurrency. The use of connection pools, as implemented in the code, helps to consequently reduce latency and system overhead, resulting in better performance and efficiency.

## VI. RESULTS ANALYSIS

There is a direct relationship between the amount of data to be transacted between the blockchain and the database (associated with latency), it refers to the more data that needs to be transacted, that is, mined and confirmed by the blockchain network, the higher the cost will be. final transaction fee for this data. In this project, as previously mentioned in other chapters, one of Ethereum's test networks, Sepolia, was used for its development. In this way, the costs associated with transactions are null and paid for by non-real monetary values. When it comes to assessing the performance of integrated systems, it's crucial to understand how efficiently data transactions are happening between blockchain and cloud databases.

In this project, we took a close look at the transaction times required for sending and retrieving data between the Ethereum blockchain and PostgreSQL databases using three different server setups. Analyzing transaction times helps us identify any bottlenecks, optimize performance, and make sure that the system meets the necessary speed requirements for real-world applications. In this project, it was decided to use a remote server. Using this remote server has several advantages, namely advanced computing resources and uninterrupted operation. The server in question contains the following specifications: Windows 11 Pro; 64-bit operating system, x64-based processor; AMD Ryzen 9 7950X 16-Core Processor (4.50GHz); RAM 128GB (127GB usable); Rom memory Samsung SSD 870 EVO 1TB. This remote server will be an essential component, as it will facilitate various tasks such as algorithm efficiency tests, latency assessment, and various other points relating to the maximum capabilities that the system supports. These will be exposed to large workloads, such as generating large amounts of data packets for the database and blockchain, the number of users within the system, and their hardware needs. So, let's dive into a detailed analysis of the transaction times we observed for the various server configurations in the following Figures.

In Fig. 5, we can observe that the transaction time increases as the data volume grows. However, the difference in transaction time across different servers is minimal, with about a 1 or 2-second difference between transactions involving 60 packages and those involving $60 \times 10^{12}$ packages. Consequently, it is difficult to determine which server version is superior in terms of transaction time for shipping.

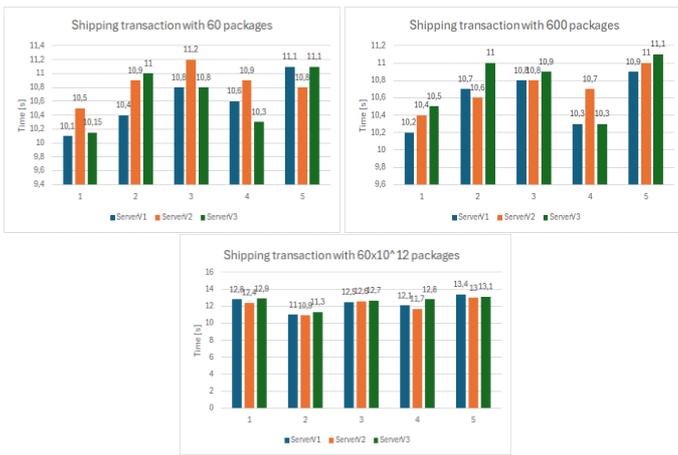

**Fig. 5 Graphical representation of shipping transaction time**

Based on Fig. 6 data, ServerV2 is the best choice for recovery transactions, as it shows consistently a recovery time of 0 seconds across all transaction sizes.

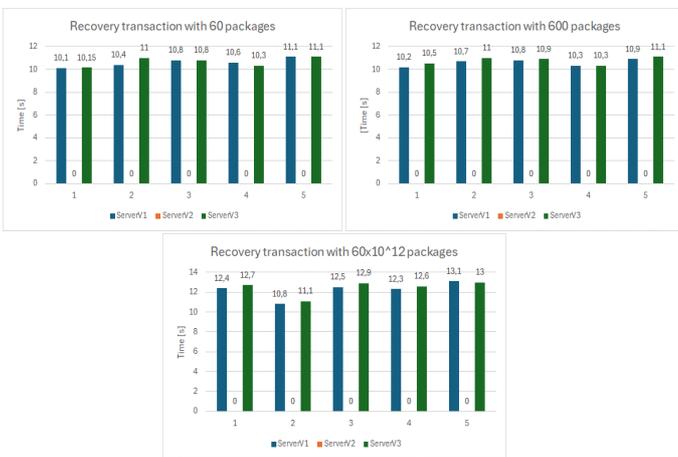

**Fig. 6 Graphical representation of recovery transaction time**

Companies today are increasingly interested in programs that prioritize energy efficiency, often referred to as "green" initiatives. Our program, designed with a strong focus on these criteria, has proven its effectiveness through previous results. These programs are carefully developed to operate with minimal consumption of computational resources, such as CPU, RAM, ROM, GPU, and others. This approach not only reduces energy usage but also aligns with the growing emphasis on environmental awareness and sustainability. By lowering energy consumption during execution, these programs make a significant contribution to conservation efforts, which are crucial in a world where energy use directly impacts greenhouse gas emissions and global warming. Furthermore, our program promotes sustainability by prolonging the lifespan of the devices it runs on. By efficiently utilizing resources, it helps minimize wear and tear on hardware components, thereby reducing the rate of device obsolescence and the generation of electronic waste. This aligns perfectly with the values of large companies that seek solutions prioritizing environmental preservation alongside operational efficiency. Moreover, our program enhances operational efficiency within organizations. By optimizing resource requirements, it improves overall system performance, resulting in time and resource savings. This dual benefit of environmental sustainability and operational efficiency makes our program an attractive choice for companies looking to reduce their carbon footprint while boosting productivity. Evaluating the sustainability and operational costs of any computing system involves considering energy consumption as a vital metric. In this project, we measured the energy consumption, including CPU, Memory, and Disk usage, of the dedicated server running our integrated blockchain and database system. By analyzing these metrics, we aim to understand how energy-efficient our solution is, pinpoint areas for improvement, and ensure that the system can operate sustainably under different workloads. So, let's dig into a comprehensive analysis of the energy consumption we observed on the dedicated server.

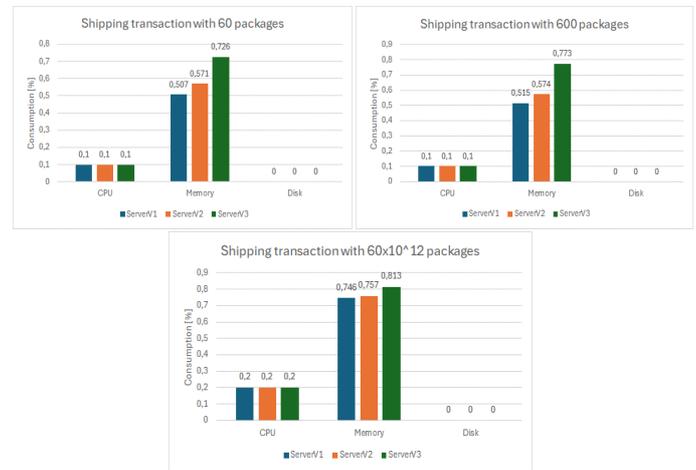

**Fig. 7 Graphical representation of shipping transaction energy consumption on a server**

As we can see in Fig. 6, serverV1 shows consistently the lowest memory consumption across all transaction sizes, with equal CPU and Disk consumption compared to the other servers. Therefore, ServerV1 is the best choice for shipping transactions when it comes to energy consumption.

When analyzing the graphic of Fig. 7, we can conclude that serverV1 is again the best choice, for recovery transactions when it comes to energy consumption.

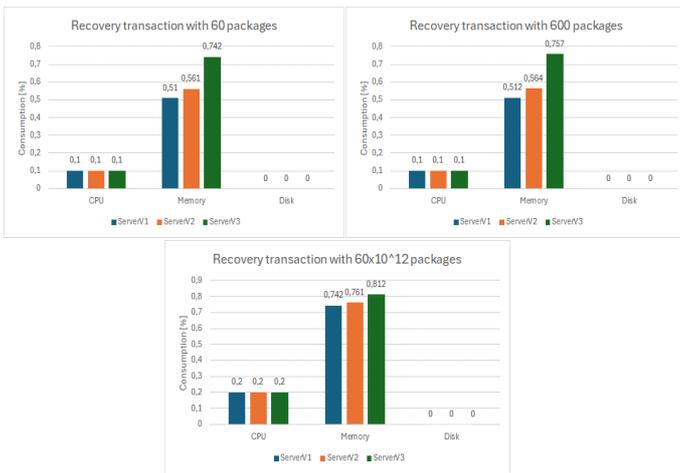

**Fig. 7 Graphical representation of recovery transaction energy consumption on a server**

In Solidity, strings can support up to $2^{256} - 1$ bytes theoretically. However, practical concerns like gas costs or the block gas limit enforced by Ethereum significantly limit efficient on-chain data storage. In Ethereum, gas prices are mostly determined by the volume of information that is placed in storage. When we store a longer string, every character is 4 bytes thus increasing the size of the data and consequently escalating the expenses.

Considering that 'x' represents the gas price per byte and 'y' represents the number of packages to be stored in a string, we can create an equation depending on the gas price and the number of packages (set of 36 characters) in a string:

- String length = 36 * 4 * y = 144y bytes.
- Gas price = 144yx.

In theory, we have a limitation of $\frac{(2^{256} - 1)}{144}$ packages in every transaction to the blockchain. However, practically this is not reliable because of the gas price, it's needed to achieve this type of transaction.

## VII. CONCLUSIONS

Throughout this project, the main objective was to address the challenges that come with data security in cloud-hosted database systems. We thus explore the potential of combining blockchain technology with cloud databases, specifically using the Ethereum blockchain network together with PostgreSQL databases. With this project we reached significant milestones in efforts to create an innovative, more secure and transparent data storage environment. The successful implementation of smart contracts, algorithmic optimizations, and seamless integration between blockchain and cloud databases played an important role in the success of this project. Our solution not only enables smooth interaction between blockchain and cloud databases, but also addresses important issues such as user privacy, authentication, authorization and data integrity. By implementing strong security measures such as data encryption and access control mechanisms, we introduce further global security reinforcements into the system. Additionally, by optimizing performance and scalability, we overcome challenges typically associated with integrating blockchain and cloud databases. Techniques such as connection pools that have greatly improved system efficiency and responsiveness, resulting in a more pleasant and reliable user experience.

In conclusion, the integration of Ethereum blockchain technology and PostgreSQL database presents an excellent opportunity as it ensures data security and integrity in distributed environments like the cloud. Despite the challenges we faced throughout the development of the project, the benefits offered by our solution, especially after optimizations, overcame all remaining obstacles, making it a viable and promising option in the current market. The results presented in the previous chapter prove that through the implementation of smart contracts and synchronization between the two environments, we achieved transparency and improved security in data storage. However, it is important to note that this project is just the first step in an ongoing journey of innovation and improvement. We recognize that there are significant opportunities to further expand and enhance these two blockchain and cloud database integration technologies.

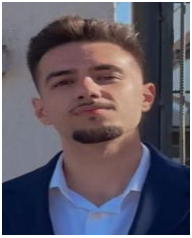

**R. C. Rodrigues** is currently studying for a bachelor's degree in electronics, telecommunications, and computers at the Instituto Superior de Engenharia de Lisboa, Lisboa–Portugal. His research interests include cybersecurity, data privacy, and distributed systems.

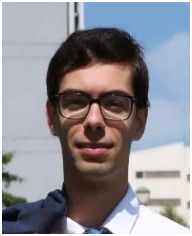

**P. M. C. Mateus** is currently studying for a bachelor's degree in electronics, telecommunications, and computers at the Instituto Superior de Engenharia de Lisboa, Lisboa–Portugal. His research interests include cybersecurity and database management systems.

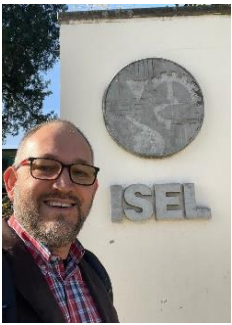

**V. R. Q. Leithardt** is Professor the Instituto Superior de Engenharia de Lisboa (ISEL), Lisboa – Portugal, and Senior Member of IEEE. He received the Ph.D. degree in Computer Science from INF-UFRGS, Brazil, in 2015. He is also an Integrated Member of the Center of Technology and Systems (UNINOVA-CTS) and Associated Lab of Intelligent Systems (LASI), 2829-516 Caparica, Portugal.